\documentclass{PoS}

\newcommand{\Slash}[1]{{\ooalign{\hfil/\hfil\crcr$#1$}}}

\title{Dirac-mode expansion for confinement and chiral symmetry breaking}

\ShortTitle{Dirac-mode expansion for confinement and chiral symmetry breaking}

\author{\speaker{Hideo Suganuma}, Shinya Gongyo, Takumi Iritani \\
        Department of Physics, Kyoto University, Kitashirakawaoiwake, Sakyo, Kyoto
606-8502, Japan\\
        E-mail: \email{suganuma@scphys.kyoto-u.ac.jp}}

\abstract{
We develop a manifestly gauge-covariant expansion and projection 
using the eigen-mode of the QCD Dirac operator 
$\Slash D=\gamma^\mu D^\mu$. 
Applying this method to the Wilson loop and the Polyakov loop, 
we perform a direct analysis of 
the correlation between confinement and chiral symmetry breaking 
in SU(3) lattice QCD calculation 
on $6^4$ at $\beta$=5.6 at the quenched level. 
Notably, 
the Wilson loop is found to obey the area law, 
and the slope parameter corresponding to the string tension or 
the confinement force is almost unchanged, 
even after removing the low-lying Dirac modes, 
which are responsible to chiral symmetry breaking.
We find also that the Polyakov loop remains to be almost zero 
even without the low-lying Dirac modes, 
which indicates the $Z_3$-unbroken confinement phase.
These results indicate that one-to-one correspondence does not 
hold between confinement and chiral symmetry breaking 
in QCD. 
}

\FullConference{
The 30 International Symposium on Lattice Field Theory - Lattice 2012,\\
		June 24-29, 2012\\
		Cairns, Australia}

\begin{document}

\section{Introduction: 
relation between confinement and chiral symmetry breaking}

Quantum chromodynamics (QCD) exhibits interesting nonperturbative phenomena 
such as color confinement and chiral symmetry breaking \cite{NJL61} 
in the low-energy region.
In particular, it is an important issue to investigate 
the correlation between confinement 
and chiral symmetry breaking \cite{SST95,M95W95,G06BGH07}. 
However, their relation is not yet clarified directly from QCD,
although the strong correlation between them has been suggested 
by the simultaneous phase transitions of deconfinement and 
chiral restoration in lattice QCD both 
at finite temperature \cite{R12} 
and in a small-volume box \cite{R12}.

The close relation between confinement and chiral symmetry breaking 
has been also suggested in terms of the monopole 
degrees of freedom \cite{SST95,M95W95}, 
which topologically appears in QCD 
by taking the maximally Abelian gauge 
\cite{N74tH81,KSW87,SNW94}. 
For example, by removing the monopoles, 
confinement and chiral symmetry breaking are 
simultaneously lost in lattice QCD \cite{M95W95}, 
as schematically shown in Fig.1. 
This indicates an important role of the monopole 
to both confinement and chiral symmetry breaking, 
and these two nonperturbative QCD phenomena seem 
to be related via the monopole.

\begin{figure}[ht]
\begin{center}
\hspace{-0.5cm} \includegraphics[scale=0.326]{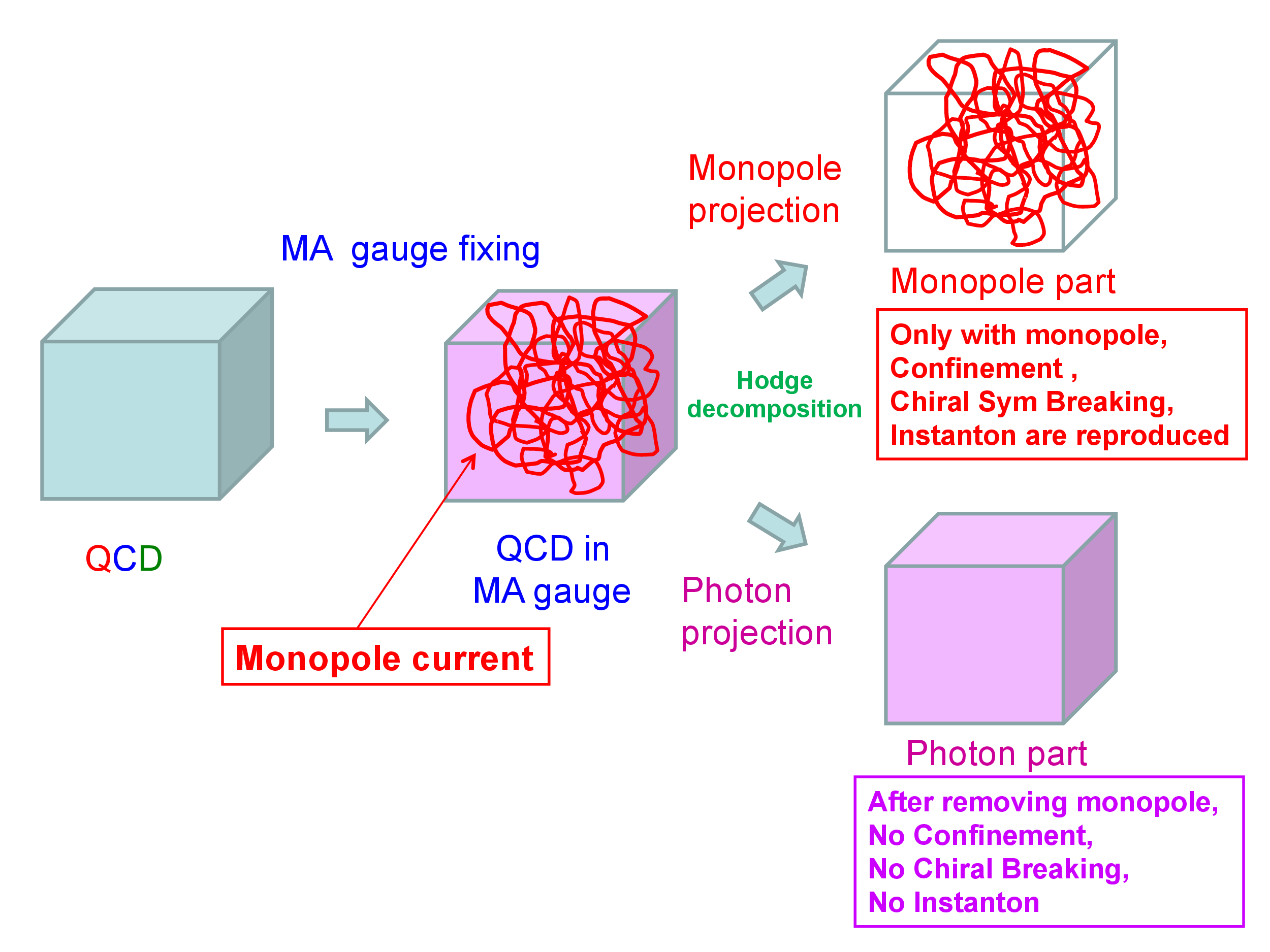}
\caption{
An illustration of 
the relevant role of monopoles to nonperturbative QCD. 
In the maximally Abelian gauge, 
QCD becomes Abelian-like due to the 
large off-diagonal gluon mass of about 1GeV \cite{AS99}, and 
there appears a global network of the monopole current \cite{KSW87,SNW94}. 
By the Hodge decomposition, the QCD system can be divided into 
the monopole part and the photon part. 
The monopole part has confinement \cite{SNW94},
chiral symmetry breaking \cite{M95W95} and instantons \cite{STSM95},
while the photon part does not have all of them.
}
\end{center}
\end{figure}

However, as a possibility, removing the monopoles 
may be ``too fatal'' for most nonperturbative properties. 
If this is the case, nonperturbative QCD 
phenomena are simultaneously lost by their cut. 

In fact, {\it if only the relevant ingredient of 
chiral symmetry breaking is carefully removed, 
how will be confinement?}
To get the answer, we perform a direct investigation between 
confinement and chiral symmetry breaking, 
using the Dirac-mode expansion and projection \cite{GIS12}.

\section{Gauge-invariant formalism of Dirac-mode expansion and projection} 

In this paper, we develop a manifestly gauge-covariant expansion/projection 
of QCD operators such as the Wilson loop and the Polyakov loop, 
using the eigen-mode of the QCD Dirac operator 
$\Slash D=\gamma^\mu D^\mu$, and investigate 
the relation between confinement and chiral symmetry breaking \cite{GIS12}. 

\subsection{Eigen-mode of Dirac operator in lattice QCD}

In lattice QCD with spacing $a$, 
the Dirac operator 
$\Slash D = \gamma_\mu D_\mu$ is expressed with $U_\mu(x)$ as
\begin{eqnarray}
      \Slash{D}_{x,y} 
      \equiv \frac{1}{2a} \sum_{\mu=1}^4 \gamma_\mu 
\left[ U_\mu(x) \delta_{x+\hat{\mu},y}
        - U_{-\mu}(x) \delta_{x-\hat{\mu},y} \right],
\end{eqnarray}
with $U_{-\mu}(x)\equiv U^\dagger_\mu(x-\hat \mu)$.
Adopting hermitian $\gamma$-matrices $\gamma_\mu^\dagger=\gamma_\mu$, 
$\Slash D$ is anti-hermitian and satisfies 
$\Slash D_{y,x}^\dagger=-\Slash D_{x,y}$.
The normalized eigen-state $|n \rangle$ 
of the Dirac operator $\Slash D$ is introduced as 
\begin{eqnarray}
\Slash D |n\rangle =i\lambda_n |n \rangle
\end{eqnarray}
with $\lambda_n \in {\bf R}$.
Because of $\{\gamma_5,\Slash D\}=0$, the state 
$\gamma_5 |n\rangle$ is also an eigen-state of $\Slash D$ with the 
eigenvalue $-i\lambda_n$. 
The Dirac eigenfunction $\psi_n(x)\equiv\langle x|n \rangle$ 
obeys $\Slash D \psi_n(x)=i\lambda_n \psi_n(x)$, 
and its explicit form of the eigenvalue equation in lattice QCD is 
\begin{eqnarray}
\frac{1}{2a} \sum_{\mu=1}^4 \gamma_\mu
[U_\mu(x)\psi_n(x+\hat \mu)-U_{-\mu}(x)\psi_n(x-\hat \mu)]
=i\lambda_n \psi_n(x).
\end{eqnarray}
The Dirac eigenfunction $\psi_n(x)$ can be 
numerically obtained in lattice QCD, besides a phase factor. 

According to 
$U_\mu(x) \rightarrow V(x) U_\mu(x) V^\dagger (x+\hat\mu)$, 
the gauge transformation of $\psi_n(x)$ is found to be 
\begin{eqnarray}
\psi_n(x)\rightarrow V(x) \psi_n(x),
\label{eq:GTprop}
\end{eqnarray}
which is the same as that of the quark field.
To be strict, for the Dirac eigenfunction, 
there can appear an irrelevant $n$-dependent global phase factor 
as $e^{i\varphi_n[V]}$,
according to the arbitrariness of the definition of $\psi_n(x)$.

From the Banks-Casher relation \cite{BC80}, 
the quark condensate $\langle\bar qq \rangle$, 
the order parameter of chiral symmetry breaking, is given by 
the zero-eigenvalue density $\rho(0)$ 
of the Dirac operator $\Slash D$:
\begin{eqnarray}
\langle \bar qq \rangle=-\lim_{m \to 0} \lim_{V \to \infty} 
\pi\rho(0),
\end{eqnarray}
where the spectral density $\rho(\lambda)$ is given by 
$\rho(\lambda)\equiv 
\frac1V\sum_{n}\langle \delta(\lambda-\lambda_n)\rangle$ 
with space-time volume $V$.
Thus, the low-lying Dirac modes can be regarded as the essential modes 
responsible to spontaneous chiral-symmetry breaking in QCD.

\subsection{Operator formalism in lattice QCD}

The recent analysis of QCD with the Fourier expansion of the gluon field 
quantitatively reveals that 
quark confinement originates from low-momentum gluons below about 1GeV 
in both Landau and Coulomb gauges \cite{YS0809}. 
This method seems powerful but accompanies some gauge dependence. 
To keep the gauge symmetry manifestly, 
we take the ``operator formalism'' in lattice QCD \cite{GIS12}.

We define the link-variable operator $\hat U_\mu$ 
by the matrix element of 
\begin{eqnarray}
\langle x |\hat U_\mu|y\rangle =U_\mu(x)\delta_{x+\hat \mu,y}.
\end{eqnarray}
The Wilson-loop operator $\hat W$ is defined as the product of 
$\hat U_\mu$ along a rectangular loop,
\begin{eqnarray}
\hat W \equiv \prod_{k=1}^N \hat U_{\mu_k}
=\hat U_{\mu_1}\hat U_{\mu_2} \cdots \hat U_{\mu_N}.
\end{eqnarray}
For arbitrary loops, one finds $\sum_{k=1}^N \hat \mu_k=0$.
We note that the functional trace of the Wilson-loop operator 
$\hat W$ is proportional to the ordinary vacuum expectation value 
$\langle W \rangle$ of the Wilson loop:
\begin{eqnarray}
{\rm Tr} \ \hat W&=&{\rm tr}\sum_x \langle x |\hat W|x \rangle
={\rm tr}\sum_x \langle x| \hat U_{\mu_1}\hat U_{\mu_2} 
\cdots \hat U_{\mu_N}|x\rangle \nonumber\\
&=& {\rm tr} \sum_{x_1, x_2, \cdots, x_N }
\langle x_1| \hat U_{\mu_1}|x_2 \rangle
\langle x_2| \hat U_{\mu_2}|x_3 \rangle
\langle x_3| \hat U_{\mu_3}|x_4 \rangle
\cdots \langle x_N|\hat U_{\mu_N}|x_1\rangle \nonumber\\
&=&{\rm tr} \sum_x 
\langle x| \hat U_{\mu_1}|x+\hat \mu_1 \rangle
\langle x+\hat \mu_1| \hat U_{\mu_2}|x+\sum_{k=1}^2\hat \mu_k \rangle
\cdots \langle x+\sum_{k=1}^{N-1}\hat \mu_k|\hat U_{\mu_N}|x\rangle \nonumber\\
&=&\sum_x {\rm tr}\{ U_{\mu_1}(x) U_{\mu_2}(x+\hat \mu_1)
U_{\mu_3}(x+\sum_{k=1}^2 \hat \mu_k)
\cdots U_{\mu_N}(x+\sum_{k=1}^{N-1} \hat \mu_k)\} 
=\langle W \rangle \cdot {\rm Tr}\ 1.
\label{eq:TrWLO}
\end{eqnarray}
Here, ``Tr'' denotes the functional trace, 
and ``tr'' the trace over SU(3) color index.

The Dirac-mode matrix element of the link-variable operator 
$\hat U_{\mu}$ can be expressed with $\psi_n(x)$:
\begin{eqnarray}
\langle m|\hat U|n \rangle=\sum_x\langle m|x \rangle 
\langle x|\hat U_{\mu}|x+\hat \mu \rangle \langle x+\hat \mu|n\rangle
=\sum_x \psi_m^\dagger(x) U_\mu(x)\psi_n(x+\hat \mu).
\end{eqnarray}
Although the total number of the matrix element is very huge, 
the matrix element is calculable and gauge invariant, 
apart from an irrelevant phase factor.
Using the gauge transformation (\ref{eq:GTprop}), we find 
the gauge transformation of the matrix element as \cite{GIS12}
\begin{eqnarray}
\langle m|\hat U_\mu|n \rangle
&=&\sum_x \psi^\dagger_m(x)U_\mu(x)\psi_n(x+\hat\mu) \nonumber\\
&\rightarrow&
\sum_x\psi^\dagger_m(x)V^\dagger(x)\cdot V(x)U_\mu(x)V^\dagger(x+\hat \mu)
\cdot V(x+\hat \mu)\psi_n(x+\hat \mu) \nonumber\\
&=&\sum_x\psi_m^\dagger(x)U_\mu(x)\psi_n(x+\hat \mu)
=\langle m|\hat U_\mu|n\rangle.
\end{eqnarray}
To be strict, there appears an $n$-dependent global phase factor, 
corresponding to the arbitrariness of the phase in the basis 
$|n \rangle$. However, this phase factor cancels 
as $e^{-i\varphi_n} e^{i\varphi_n}=1$ 
between $|n \rangle$ and $\langle n |$, and does not appear 
for QCD physical quantities including the Wilson loop.

\subsection{Dirac-mode expansion and projection}

From the completeness of the Dirac-mode basis, 
$\sum_n|n\rangle \langle n|=1$, 
arbitrary operator $\hat O$ can be expanded 
in terms of the Dirac-mode basis $|n \rangle$ as
\begin{eqnarray}
\hat O=\sum_n \sum_m |n \rangle \langle n|\hat O|m \rangle \langle m|, 
\label{eq:dirac-mode-expansion}
\end{eqnarray}
which is the theoretical basis of 
the Dirac-mode expansion \cite{GIS12}. 
Note here that this procedure is just the insertion 
of unity, and is of course mathematically correct.

Based on this relation, the Dirac-mode expansion and projection 
can be defined. We define the projection operator $\hat P$ 
which restricts the Dirac-mode space, 
\begin{eqnarray}
\hat P\equiv \sum_{n \in A}|n\rangle \langle n|,
\end{eqnarray} 
where $A$ denotes arbitrary set of Dirac modes. 
In $\hat P$, the arbitrary phase cancels 
between $|n\rangle$ and $\langle n|$. 
One finds $\hat P^2=\hat P$ and $\hat P^\dagger =\hat P$.
The typical projections are 
IR-cut and UV-cut of the Dirac modes:
\begin{eqnarray}
\hat P_{\rm \ IR} \equiv 
\sum_{|\lambda_n| \ge \Lambda_{\rm IR}}|n \rangle \langle n|,
\qquad 
\hat P_{\rm \ UV} \equiv 
\sum_{|\lambda_n| \le \Lambda_{\rm UV}}|n \rangle \langle n|.
\end{eqnarray} 

Using the projection operator $\hat P$, we define 
the Dirac-mode projected link-variable operator, 
\begin{eqnarray}
\hat U^P_\mu \equiv \hat P \hat U_\mu \hat P
=\sum_{m \in A}\sum_{n \in A} 
|m\rangle \langle m|\hat U_\mu|n\rangle \langle n|.
\end{eqnarray}
During this projection, there appears some nonlocality in general, 
but it would not be important for the argument of 
large-distance properties such as confinement. 
From the Wilson-loop operator 
$\hat W \equiv \prod_{k=1}^N\hat U_{\mu_k}$, 
we define the Dirac-mode projected Wilson-loop operator 
$\hat W^P \equiv \prod_{k=1}^N \hat U^P_{\mu_k}$, 
and rewrite its functional trace in terms of the Dirac basis as \cite{GIS12}
\begin{eqnarray}
{\rm Tr} \ \hat W^P &=& {\rm Tr} \ \prod_{k=1}^N \hat U^P_{\mu_k}
={\rm Tr} \ \hat U^P_{\mu_1}\hat U^P_{\mu_2}\cdots \hat U^P_{\mu_N} 
={\rm Tr} \ \hat P \hat U_{\mu_1} \hat P \hat U_{\mu_2} \hat P 
\cdots \hat P \hat U_{\mu_N} \hat P \nonumber\\
&=&{\rm tr} \sum_{n_1, n_2, \cdots, n_N \in A} 
\langle n_1| \hat U_{\mu_1} |n_2 \rangle 
\langle n_2| \hat U_{\mu_2} |n_3 \rangle \cdots
\langle n_N| \hat U_{\mu_N}|n_{1} \rangle,
\end{eqnarray}
which is manifestly gauge invariant. 
Here, the arbitrary phase factor 
cancels between $|n_k \rangle$ and $\langle n_k|$. 
Its gauge invariance is also numerically checked 
in the lattice QCD Monte Carlo calculation.

From ${\rm Tr} \ \hat W^P(R,T)$ on the $R \times T$ rectangular loop, 
we define Dirac-mode projected potential, 
\begin{eqnarray}
V^P(R)\equiv -\lim_{T \to \infty} \frac{1}{T}
{\rm ln} \{{\rm Tr} \ \hat W^P(R,T)\}.
\end{eqnarray}

On a periodic lattice of $V =L^3 \times N_t$, 
we define the Dirac-mode projected Polyakov loop \cite{GIS12}: 
\begin{eqnarray}
\langle L_{P}^{\rm proj.} \rangle\equiv 
\frac{1}{3V} {\rm Tr} \ \prod_{i=1}^{N_t} \hat{U}_4^P=\frac{1}{3V} 
{\rm Tr} \ (\hat{U}_4^P)^{N_t}=\frac{1}{3V}{\rm tr} 
\sum_{n_1,.., n_{N_t} \in A}
   \langle n_1 | \hat{U}_4 | n_2 \rangle \langle n_2 | \hat{U}_4 | n_3 \rangle
      \cdots \langle n_{N_t} | \hat{U}_4 | n_1 \rangle,~~~~~~~ 
  \end{eqnarray}
which is also manifestly gauge-invariant.

\section{Analysis of confinement in terms of Dirac modes in QCD}

In this paper, we mainly consider the removal of low-lying Dirac modes, 
i.e., the IR-cut case. 
Using the Dirac-mode expansion and projection method, 
we calculate the IR-Dirac-mode-cut Wilson loop ${\rm Tr}~W^P(R,T)$, 
the IR-cut inter-quark potential $V^P(R)$, 
and the IR-Dirac-mode-cut Polyakov loop $\langle L_P \rangle_{\rm IR}$ 
in a gauge-invariant manner \cite{GIS12}. 
Here, we can directly investigate 
the relation between chiral symmetry breaking 
and confinement as the area-law behavior of the Wilson loop, 
since the low-lying Dirac modes are responsible to 
chiral symmetry breaking.

As a technical difficulty, we have to deal with 
huge dimensional matrices and their products.  
Actually, the total matrix dimension of 
$\langle m|\hat U_\mu|n\rangle$ is (Dirac-mode number)$^2$. 
On the $L^4$ lattice, the Dirac-mode number is 
$L^4 \times N_c \times$  4, 
which can be reduced to be $L^4 \times N_c$, 
using the Kogut-Susskind technique \cite{R12}.
%
Even for the projected operator, where the Dirac-mode space is 
restricted, the matrix is generally still huge. 
At present, we use a small-size lattice 
in the actual lattice QCD calculation.

We use SU(3) lattice QCD with the standard plaquette action 
at $\beta=5.6$ 
(i.e., $a\simeq 0.25{\rm fm}$) on $6^4$ at the quenched level.
The periodic boundary condition is imposed for the gauge field.
We show in Fig.2(a) the spectral density $\rho(\lambda)$ of 
the QCD Dirac operator $\Slash D$. 
The chiral property of $\Slash D$ leads to 
$\rho(-\lambda)=\rho(\lambda)$.
Figure~2(b) is the IR-cut Dirac spectral density 
$\rho_{\rm IR}(\lambda)\equiv 
\rho(\lambda)\theta(|\lambda|-\Lambda_{\rm IR})$ 
with the IR-cutoff $\Lambda_{\rm IR}=0.5a^{-1}\simeq 0.4{\rm GeV}$.
By removing the low-lying Dirac modes, 
the chiral condensate is largely reduced as 
$\langle \bar qq\rangle_{\Lambda_{\rm IR}}/
\langle \bar qq\rangle \simeq 0.02$
around the physical region of $m_q \simeq 5{\rm MeV}$.

\begin{figure}[h]
\begin{center}
\includegraphics[scale=0.4]{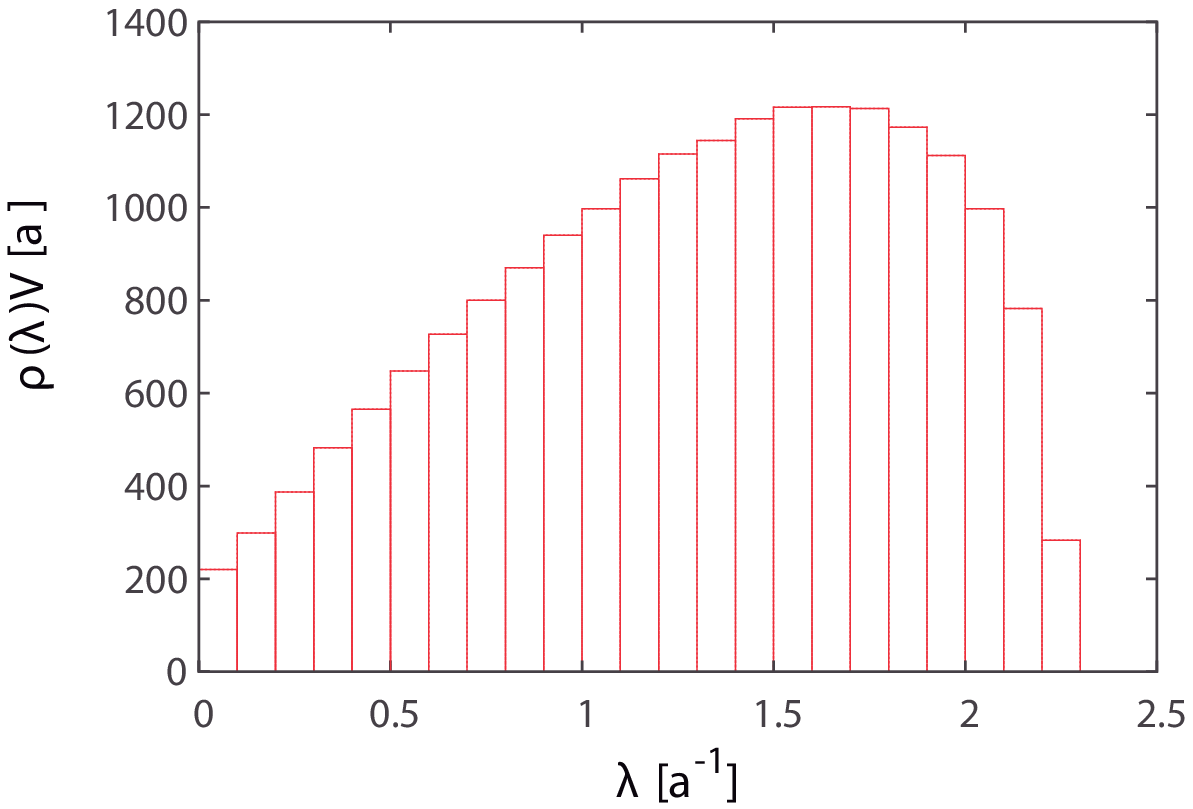}
\hspace{1cm}
\includegraphics[scale=0.4]{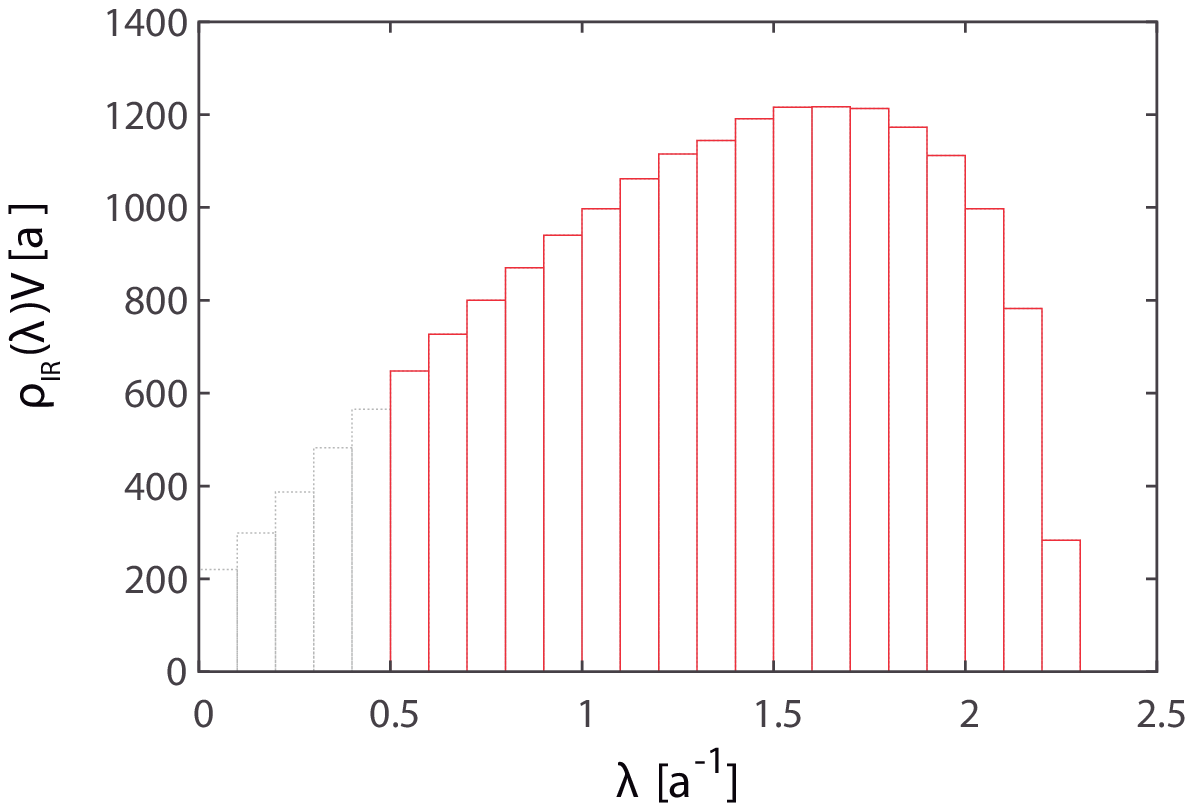}
\caption{
(a) The Dirac spectral density $\rho(\lambda)$ 
in lattice QCD at $\beta$=5.6 and $6^4$.
The volume $V$ is multiplied.
(b) The IR-cut Dirac spectral density 
$\rho_{\rm IR}(\lambda)\equiv 
\rho(\lambda)\theta(|\lambda|-\Lambda_{\rm IR})$ 
with the IR-cutoff $\Lambda_{\rm IR}=0.5a^{-1}\simeq 0.4{\rm GeV}$.
}
\end{center}
\end{figure}

Figure~3 shows the IR-Dirac-mode-cut 
Wilson loop $\langle W^P(R,T) \rangle \equiv {\rm Tr} \hat W^P(R,T)$, 
the IR-cut inter-quark potential $V^P(R)$, and 
the IR-Dirac-mode-cut Polyakov loop $\langle L_P \rangle_{\rm IR}$, 
after the removal of the low-lying Dirac modes. 
These Dirac-mode projected quantities are obtained 
in lattice QCD with the IR-cut of 
$\rho_{\rm IR}(\lambda)\equiv 
\rho(\lambda)\theta(|\lambda|-\Lambda_{\rm IR})$ 
with the IR-cutoff $\Lambda_{\rm IR}=0.5a^{-1}\simeq 0.4{\rm GeV}$.

\begin{figure}[h]
\begin{center}
\includegraphics[scale=0.4]{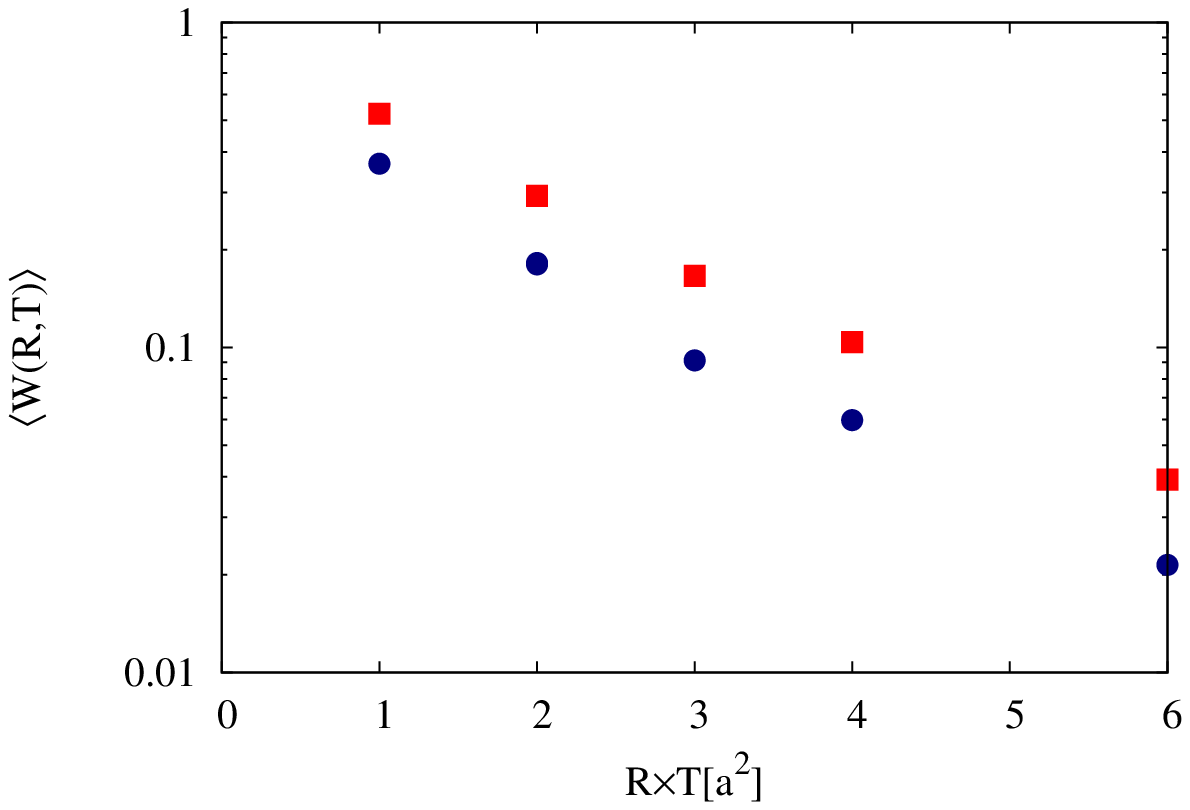}
\includegraphics[scale=0.4]{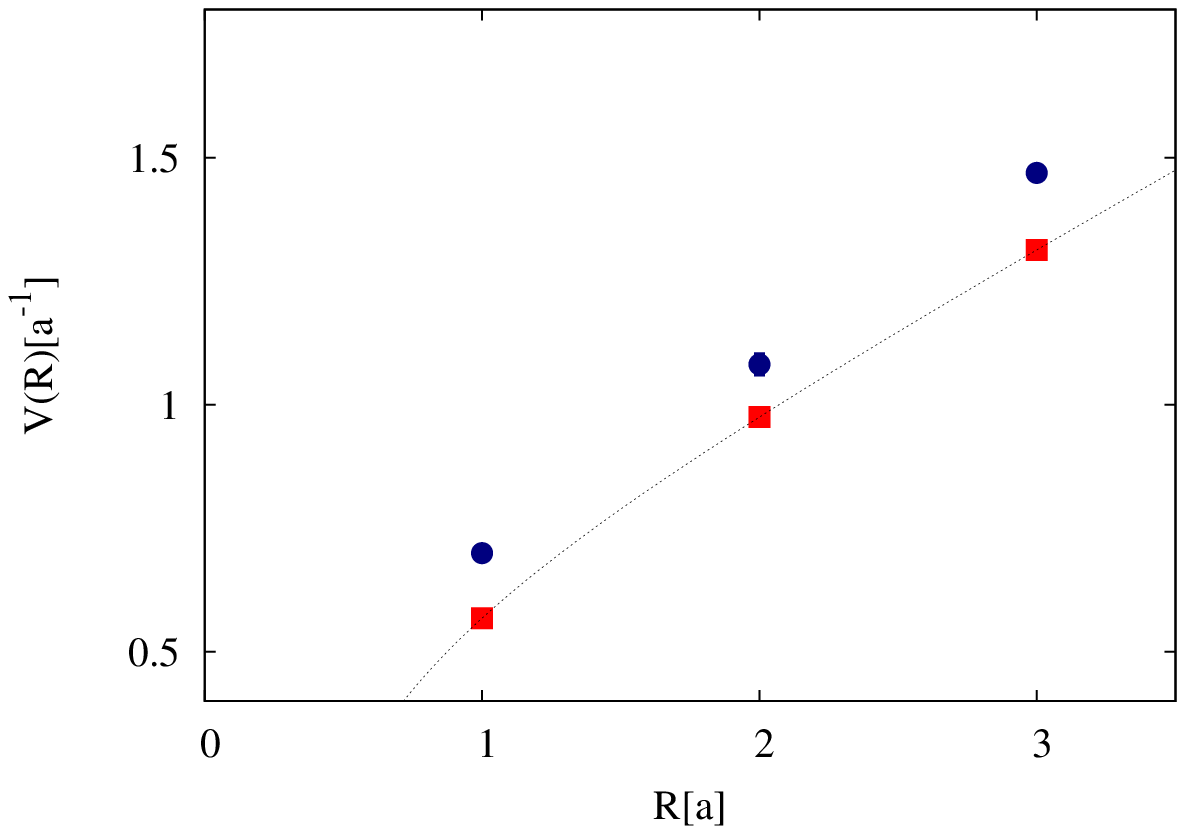}
\includegraphics[scale=0.4]{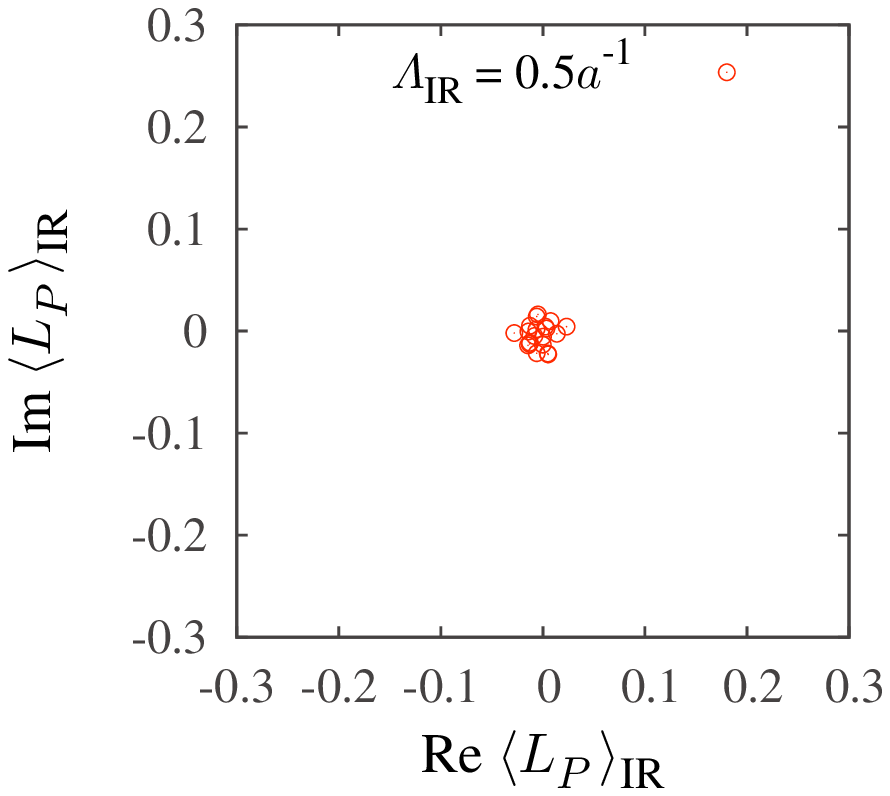}
\caption{
The attice QCD results  
after the removal of low-lying Dirac modes \cite{GIS12}, 
which gives 
$\rho_{\rm IR}(\lambda)\equiv 
\rho(\lambda)\theta(|\lambda|-\Lambda_{\rm IR})$ 
with the IR-cutoff $\Lambda_{\rm IR}=0.5a^{-1} \simeq 0.4{\rm GeV}$.
(a) The IR-cut Wilson loop ${\rm Tr}~W^P(R,T)$ (circle) 
after removing the IR Dirac modes, plotted against $R \times T$. 
The slope parameter $\sigma^P$ is almost the same as that of 
the original Wilson loop (square).
(b) The IR-cut inter-quark potential (circle), 
which is almost unchanged from the original one (square),
apart from an irrelevant constant.
(c) The scatter plot of the IR-Dirac-mode-cut Polyakov loop 
$\langle L_P \rangle_{\rm IR}$: 
its zero-value indicates $Z_3$-unbroken confinement phase.
}
\end{center}
\end{figure}

Remarkably, 
even after removing the coupling to the low-lying Dirac modes, 
which are responsible to chiral symmetry breaking, 
the IR-Dirac-mode-cut  Wilson loop is found to 
obey the area law as 
$\langle W^P(R,T)\rangle \propto e^{-\sigma^P RT}$, 
and the slope parameter $\sigma^P$ 
corresponding to the string tension 
or the confinement force is almost unchanged as 
$\sigma^P \simeq \sigma$.
Accordingly, as shown in Fig.3(b), 
the IR-cut inter-quark potential $V^P(R)$ 
is almost unchanged from the original one, 
apart from an irrelevant constant. 
Also from Fig.3(c), we find that the IR-Dirac-mode-cut 
Polyakov loop is almost zero, i.e., $\langle L_P \rangle_{\rm IR} \simeq 0$, 
which indicates $Z_3$-unbroken confinement phase.
In fact, quark confinement is kept 
in the absence of the low-lying Dirac modes or 
the essence of chiral symmetry breaking \cite{GIS12}. 
This result seems consistent with 
Gattringer's formula \cite{G06BGH07} and Lang-Schrock's result \cite{LS11}.

We also investigate the UV-cut of Dirac modes in lattice QCD, 
and find that the confining force is almost unchanged 
by the UV-cut \cite{GIS12}, which seems consistent with 
the lattice result of Synatschke-Wipf-Langfeld \cite{SWL08}.
Furthermore, we examine ``intermediate-cut'' of Dirac modes, 
and obtain almost the same confining force \cite{GIS12}. 
Then, we conjecture that the ``seed'' of confinement is 
distributed not only in low-lying Dirac modes but also 
in a wider region of the Dirac-mode space. 

Our lattice QCD results suggest some independence 
between chiral symmetry breaking and color confinement, 
which may lead to richer phase structure in QCD. 
For example, the phase transition point can be different 
between deconfinement and chiral restoration 
in the presence of strong electro-magnetic fields, 
because of their nontrivial effect on chiral symmetry \cite{ST9193}.

\section*{Acknowledgements}
The lattice QCD calculation has been done on 
NEC-SX8R and NEC-SX9 at Osaka University.
H.S. is supported by the Grant for Scientific Research 
[(C) No.~23540306, Priority Areas ``New Hadrons'' (E01:21105006)], 
and S.G. and T.I. are supported by a Grant-in-Aid for JSPS Fellows
 [No.23-752, 24-1458] 
from the Ministry of Education, Culture, Science and Technology 
of Japan.

\end{document}